\documentstyle[12pt]{article}

\begin{document}
\begin{flushright}
{\large \bf UCL-IPT-98-02}
\end{flushright}
\vskip 5 mm
\begin{center}
{\Large \bf
A Model for a Dynamically Induced Large \\
 $CP$-Violating Phase}\footnote{This work was done in collaboration with J.-M G\'erard, R.Gonzalez Felipe and J.Weyers.}\footnote{talk presented at the VI Workshop on Particles and Fields (21-27 November 1997 in Morelia Mich., Mexico)}
\vskip 5 mm

{\large D. Del\'epine}\\
\vskip 5 mm
{\em Institut de Physique Th\'eorique, Universit\'e catholique de
Louvain,}\\

{\em  B-1348 Louvain-la-Neuve, Belgium}

\end{center}

\vskip 5 mm

\begin{abstract}
Assuming a new interaction with a  $\theta$ term for the third generation of quarks, a value of  $\theta $ different from zero dynamically induces the top-bottom mass splitting and a large CP-violating phase.

\end{abstract}

In the electroweak standard model, there are 2 independent sources of CP violation: one is the phase of the quark mixing matrix (the Cabibbo-Kobayashi-Maskawa matrix ($V_{CKM}$))\cite{kobayashi} and the second is the strong $\theta$ angle of QCD\cite{peccei}. 
\

 Nowadays, the mechanism for CP violation is still not yet understood and the CP puzzles have been turned into questions concerning tiny quark masses and a large CP-violating phase.
\
 The fact that the top quark is very heavy compared to the other quarks,  $m_{t}=180\pm 12$ GeV \cite{PDG}, suggests that the third
generation may be playing a special role in the dynamics at the electroweak
scale. So we assume the existence of a new interaction playing only with the third generation and strong enough to lead to the formation of quark-antiquark bound states which trigger dynamically the breaking of the electroweak symmetry\cite{NJL,miransky,bardeen}. This new interaction is conserving isospin symmetry between top and bottom quarks and generates a $\theta$ term.

We show that in this model, the  $\theta $ term breaks the symmetry between top and bottom as expected from general theorems \cite{vafa} and induces naturally a large CP-violating phase in $V_{CKM}$($\delta_{CKM}$) due to the smallness of the $m_{b}/m_{t}$ mass ratio.

\section{The model}

We consider a standard model Higgs sector in combination with an effective
new strong interaction acting on the third generation of quarks and
characterized by a $\theta$ term \cite{buchalla,delepine2}.

The total effective Lagrangian of our model is thus given by 

\begin{equation}
L=L_{H}+L_{\Sigma }+L_{\theta }\;,  \label{8}
\end{equation}
with $L_{H}$,$L_{\Sigma}$ and $L_{\theta}$ defined as follows:
\begin{equation}
L_{H}=D_{\mu }H^{\dagger }D^{\mu }H-m_{H}^{2}H^{\dagger }H+\left( h_{t}\bar{%
\psi}_{L}t_{R}H+h_{b}\bar{\psi}_{L}b_{R}\tilde{H}\ +\mbox{ h.c.}\right) ,
\label{1}
\end{equation}
\smallskip where $H=\left( 
\begin{array}{l}
H^{0} \\ 
H^{-}
\end{array}
\right) $ , $\tilde{H}=\left( 
\begin{array}{l}
\;H^{+} \\ 
-H^{0^{*}}
\end{array}
\right) $ and $\psi _{L}=\left( 
\begin{array}{l}
t_{L} \\ 
b_{L}
\end{array}
\right) $; $h_{t}$ and $h_{b}$ are the Yukawa couplings and $D_{\mu }$ is
the usual covariant derivative of the standard model.
$L_{\Sigma}$ parametrizes the effects of the new interaction on the top and bottom quark.

\begin{equation}
L_{\Sigma }=D_{\mu }\Sigma _{t}^{\dagger }D^{\mu }\Sigma _{t}+D_{\mu }\Sigma
_{b}^{\dagger }D^{\mu }\Sigma _{b}-m^{2}(\Sigma _{t}^{\dagger }\Sigma
_{t}+\Sigma _{b}^{\dagger }\Sigma _{b})+g(\bar{\psi}_{L}t_{R}\Sigma _{t}+%
\bar{\psi}_{L}b_{R}\tilde{\Sigma}_{b}+\mbox{h.c.})\ .  \label{3}
\end{equation}
where $\Sigma_{t}$ and $\Sigma_{b}$ are 2 complex doublet scalar fields describing the $q \bar{q}$ bound states.

\begin{equation}
\Sigma _{t}=\left( 
\begin{array}{l}
\Sigma _{t}^{0} \\ 
\Sigma _{t}^{-}
\end{array}
\right) \sim t_{R}\bar{\psi}_{L}\;,\qquad \tilde{\Sigma}_{b}=\left( 
\begin{array}{l}
\;\Sigma _{b}^{+} \\ 
-\Sigma _{b}^{0^{*}}
\end{array}
\right) \sim b_{R}\bar{\psi}_{L}  \label{2}
\end{equation}
 For the $\theta$ term, we shall take, in analogy with QCD, the lagrangian form

\begin{equation}
L_{\theta }=-\frac{\alpha }{4}\left[ i\mbox{Tr }\left( \ln U-\ln U^{\dagger
}\right) +2\theta \right] ^{2},  \label{7}
\end{equation}
with \begin{equation}
U=\left( 
\begin{array}{ll}
\Sigma _{t}^{0} & \;\Sigma _{b}^{-} \\ 
\Sigma _{t}^{+} & -\Sigma _{b}^{0^{*}}
\end{array}
\right) .  \label{6}
\end{equation}
This term typically arises as a leading term in a $1/N$ - expansion.

From Eqs.(\ref{1}) and (\ref{3}) it follows that the top and bottom
field-dependent masses are given at the tree level by the linear 
combinations

\begin{equation}
M_{t}=h_{t}H^{0}+g\Sigma _{t}^{0},\qquad M_{b}=h_{b}\tilde{H}^{0}+g\tilde{%
\Sigma}_{b}^{0},  \label{4}
\end{equation}

\section{Electroweak and isospin symmetry breakings}

 Without loss of generality, we take
the phase of the neutral Higgs field $H^{0}$ to be zero. This can always be
achieved by performing a suitable electroweak gauge transformation. We write
the VEVs of the neutral components of the fields in the form

\begin{equation}
\left\langle H^{0}\right\rangle =\frac{v}{\sqrt{2}}\;,\quad \left\langle
\Sigma _{t}^{0}\right\rangle =\frac{\sigma _{t}}{\sqrt{2}}e^{i\varphi
_{t}},\quad \left\langle \Sigma _{b}^{0}\right\rangle =\frac{\sigma _{b}}{%
\sqrt{2}}e^{i\varphi _{b}}.  \label{9}
\end{equation}

Including the radiative corrections (induced by top and bottom quark loops)\cite{fatelo},
the effective potential in terms of these VEVs reads

\begin{equation}
V=m_{H}^{2}\frac{v^{2}}{2}+\frac{m^{2}}{2}(\sigma _{t}^{2}+\sigma
_{b}^{2})-\beta \left( \mu _{t}^{2}+\mu _{b}^{2}\right) +\lambda \left( \mu
_{t}^{4}+\mu _{b}^{4}\right) +\alpha \left( \theta -\varphi _{t}+\varphi
_{b}\right) ^{2}\;,  \label{10}
\end{equation}
\newline
\noindent where
\begin{eqnarray}
\mu _{t,b}^{2} &=&|\left\langle M_{t,b}\right\rangle |^{2}=\frac{1}{2}\left(
h_{t,b}^{2}v^{2}+g^{2}\sigma _{t,b}^{2}+2h_{t,b}vg\sigma _{t,b}\cos \varphi
_{t,b}\right) ,  \label{11}
\end{eqnarray}
while $\beta $ and $\lambda $ are some effective quadratic and quartic 
couplings. In what follows we shall assume all couplings and parameters in 
the potential to be real and positive.

Note that the potential which leads to Eq.(\ref{10}) can be viewed
either as an effective renormalizable interaction or as an expansion up to
quartic terms in a cut-off theory.

The extrema conditions $\frac{\partial V}{\partial v}=\frac{\partial V}{%
\partial \sigma _{t}}=\frac{\partial V}{\partial \sigma _{b}}=\frac{\partial
V}{\partial \varphi _{t}}=\frac{\partial V}{\partial \varphi _{b}}=0$ imply
a system of equations which can be solved in a simple analytical way assuming $h_{t}=h_{b}$, $\alpha \gg \beta m_{t}^{2}$ with $m_{t}$ the physical mass of the top quark, $ \beta \gg 2 \lambda m_{t}^{2}$. In that case, the presence of a phase $\theta$ close to $\frac{\pi}{2}$ induces both isospin breaking and CP violation with \footnote{ the question of the eigenvalues of the scalar mass matrix is discussed in Ref.\cite{delepine2}}

\begin{equation}
\sigma _{b}\ll \sigma _{t}\neq 0\;,\;v\neq 0\;,\;\varphi _{t}\simeq \sigma
_{b}/\sigma _{t}\;,\;\varphi _{b}\simeq -\pi /2+\sigma _{b}/\sigma _{t}\;.
\label{22}
\end{equation}

\section{$CP$ violation}

 Let us now investigate whether this
new source of $CP$ violation can be responsible for what is observed in 
the $K^{0}-\bar{K}^{0}$ system.
Let us consider the $3\times 3$ quark mass matrices
\begin{eqnarray}
M_{u,d} &=&(h)_{u,d}v+\left( 
\begin{array}{lll}
0 &  &  \\ 
& 0 &  \\ 
&  & 1
\end{array}
\right) g\sigma _{t,b}e^{\pm i\varphi _{t,b}},  \nonumber \\
&&  \label{28} 
\end{eqnarray}
with ($h)_{u,d}$ arbitrary real matrices.

If we neglect $O(h^{2}v^{2})$ terms, $(MM^{\dagger })_{u,d}$ are
diagonalized by the following unitary matrices:

\begin{equation}
U_{u}\simeq R_{u}\left( 
\begin{array}{lll}
1 &  &  \\ 
& 1 &  \\ 
&  & e^{-i\varphi _{t}}
\end{array}
\right) \;,\;U_{d}\simeq R_{d}\left( 
\begin{array}{lll}
1 &  &  \\ 
& 1 &  \\ 
&  & e^{i\varphi _{b}}
\end{array}
\right) \;,  \label{29}
\end{equation}
with $R_{u,d}$ orthogonal. In this approximation and using  the
Cabibbo-Kobayashi-Maskawa parametrization for the quark mixing matrix \cite{kobayashi}, we get, last but not least,$\delta _{KM}\simeq -(\varphi _{t}+\varphi _{b})\simeq \pi /2 $.
In the nowadays standard parametrization of the Cabibbo-Kobayashi-Maskawa mixing matrix\cite{PDG}, the phenomenology in $K^{0}-\bar{K^{0}}$ physics requires that the  CP-violating phase   ($\delta_{13}$) be around $\frac{\pi}{2}$. The CP-violating phase in the 2 parametrizations are related by $\sin \delta _{13}\simeq \frac{\vartheta _{2}}{\vartheta _{23}}\sin \delta_{KM}$, in the small mixing angles approximation\cite{fritzsch}. From $|V_{ij}|_{KM}=|V_{ij}|_{standard}$, the experimental constraint on the ratio
\begin{equation}
 \frac{V_{ub}}{V_{cb}}=0.08\pm 0.02 
\end{equation}
require a texture for the $h$-matrices such that the mixing angles $\vartheta _{2}>\vartheta
_{3}.$

Therefore, we conclude that our solution leads indeed to a sizeable 
$CP$-violating phase
\begin{equation}
\delta _{13}\simeq \pi /2\;,  \label{39}
\end{equation}
which is welcome by phenomenology in $K^{0}-\bar{K}^{0}$ physics \cite{buras}.

\medskip

\end{document}